\documentclass[aps,prl,twocolumn,floatfix,showpacs]{revtex4}
\usepackage{epsf,amssymb,verbatim}
\newcommand{\be}{\begin{equation}}
\newcommand{\ee}{\end{equation}}
\begin{document}
\title{Load distribution in weighted complex networks}
\author{K.-I. Goh$^1$, J.~D. Noh$^2$, B. Kahng$^1$, and D. Kim$^1$}
\affiliation{
$^1$School of Physics and Center for Theoretical Physics,\\ 
Seoul National University, Seoul 151-747, Korea\\
$^2$Department of Physics, Chungnam National University, Daejon
305-764, Korea }
\date{\today}
\begin{abstract}
We study the load distribution in weighted networks by measuring 
the effective number of optimal paths passing through a given 
vertex. The optimal path, along which the total cost is minimum, 
crucially depend on the cost distribution function $p_c(c)$. 
In the strong disorder limit, where $p_c(c)\sim c^{-1}$, 
the load distribution follows a power law both in the Erd\H{o}s-R\'enyi
(ER) random graphs and in the scale-free (SF) networks, and its
characteristics are determined by the structure of the minimum 
spanning tree.
The distribution of loads at vertices with a given vertex degree 
also follows the SF nature similar to the whole 
load distribution, implying that the global transport property
is not correlated to the local structural information.
Finally, we measure the effect of disorder by the correlation 
coefficient between vertex degree and load, 
finding that it is larger for ER networks than 
for SF networks.
\end{abstract}
 \maketitle

Study of complex systems in the framework of the network representation
has attracted considerable attention as an interdisciplinary
subject~\cite{rmp,portobook,siam,vespigbook}. Of particular
interest is the emerging pattern, a power-law behavior in the
degree distribution, $p_{d}(k)\sim k^{-\gamma},$ where the degree $k$
is the number of edges connecting a given vertex. Such complex
networks are called scale-free (SF) networks~\cite{ba}. 
Transport phenomena on SF networks such as data packet transport
on the communication 
network~\cite{erramilli,takayasu,geoff,tadic_04_1,tadic_04_2,guimera}, 
random walks~\cite{noh_04} and the information exchange in social 
networks~\cite{social}, are of vital importance in both theoretical 
and practical perspectives. 
As the first step, the transport property can be studied through the quantity 
called the load introduced recently \cite{load}, or the betweenness 
centrality in social network literature~\cite{freeman}. 
To be specific, a packet leaves and arrives between a pair of 
vertices, travelling along the shortest pathway(s) between the pair. 
When the shortest pathways branch, the packet is assumed to be 
divided evenly. Then, the load $\ell_i$ of a vertex $i$ is defined as
the accumulated sum of the amount of packets passing through that
vertex when every pair of vertices send and receive a unit packet.
The load thus quantifies the level of burden of vertices
in the shortest path-based transport processes. 
The load distribution of SF networks also follows a power law, 
$p_{L}(\ell)\sim \ell^{-\delta}$, with the load exponent
$\delta$ \cite{load}. 
When packets travel with constant velocity, 
the numerics indicate that the effect of time delay on 
the load does not have effect on the shape 
of the load distribution. Therefore the load is usually
measured as the effective 
number of pathways passing through a given vertex. 
So far, the power-law behavior of the load distribution was observed 
only on binary networks, where the strength of each edge is either 1 
(present) or 0 (absent) \cite{load}.

To describe transport phenomena in a more realistic way, one has to 
take into account of the heterogeneity of elements, e.g.,
buffer sizes and/or bandwidths of each router or optical
cable. For example, the Abilene network consists of high-bandwidth backbone, 
while their sub-connecting systems do of low-bandwidth~\cite{willinger}.  
In such {\it weighted} networks,
the notion of {\it the shortest path}, the path with minimal
number of hops between two vertices in the binary network, may not be
as appropriate as the so-called {\it optimal path}, the path over
which the sum of costs becomes minimal. Thus it is natural to
generalize the load to that based on the optimal paths in weighted
networks. In general, the optimal paths in weighted networks
often take a detour with respect to the shortest path to
reduce the total cost. Such a detour makes the optimal path longer
than the shortest path~\cite{noh02,braunstein}, which 
we expect leads to redistribution of the load, the characteristics 
of which depends on the cost distribution function.
Here we will concentrate on the disorder in edges only, i.e., the
case where edges carry their own non-uniform costs with a distribution
$p_c(c)$.
In a recent paper, Park et al.~\cite{kpark} studied a related
problem based on the vertex cost, focusing on the relation
between the cost and the load of a vertex, finding that the load
of a vertex decreases exponentially with the vertex cost but
scales as a power law with respect to the degree.

To study the effect of the weight in a systematic way, we consider
a weighted network model by assigning random cost on each edge of
the static model~\cite{load}. The static model network is composed
of $N$ vertices indexed by the integers $i=1,2,\dots,N$
and $L$ edges that are added one by one avoiding the multiple connections:
Each step, an edge between a vertex pair $(i,j)$ 
is chosen with the probability $p_i p_j$ where
$p_i=i^{-\alpha}/\sum_{j}j^{-\alpha}$ and added unless it already exists. 
$\alpha$ is a control
parameter in the range $[0,1)$ and we use $L=2N$ in this work. The
network thereby constructed is a SF network with the degree exponent
$\gamma=1+1/\alpha$. Note that the $\alpha=0$ case corresponds to
the random graph of Erd\H{o}s and R\'enyi~\cite{er}. Next, cost of
each edge is assigned randomly with a given distribution function,
independent of the degrees of the vertices located at its ends.

Partly motivated by the recent observation on the real-world
weighted networks \cite{barrat03,li-chen}, we consider a family of
the cost distribution function in a power-law form $p_c(c)\sim
c^{-\omega}$, where when $\omega > 1$ ($\omega < 1$), $c$ is chosen 
as $c >1$ ($0< c < 1$). The limit $\omega\to\infty$ corresponds to the
exponential cost distribution, and the limit $\omega\to0$
does to the uniform cost distribution. As $\omega$ is
lowered from $\infty$, the heterogeneity of the cost distribution
increases and we expect that the effect of disorder would
increase, that is, longer optimal path length, $d_{\rm opt}$. The 
disorder effect
would be maximal at $\omega=1$, which corresponds to the so-called
strong disorder limit \cite{braunstein,cieplak} where the
total cost is dominated by the maximum cost over the path.
Note that if the costs are assigned as $\{\,c_i\,|\,c_1<
c_2<\dots< c_L\}$ such that $\sum_{j=1}^n c_j < c_{n+1}$, 
for instance, $c_n=2^{n-1}$, then one would get the strong 
disorder limit and in this case we have $\omega=1$.
As $\omega$ is lowered further from $\omega=1$, the effect of
disorder would decrease again. This expectation is confirmed by
numerical simulations as shown in Fig.~1(a) specifically for the SF
network with $\gamma=3$, where the ratio of the average optimal
path length and the average shortest path length $d_{\rm SP}$ are shown.
For $\omega\neq1$, the network is in the
weak disorder regime in the terminology of Ref.~\cite{braunstein}.
It is worthwhile to note that the disorder effect does not vanish
qualitatively even for the uniform distribution, the $\omega\to0$ limit.
It is not obvious if the crossover from $\omega=1^+$ to $\omega=1^-$
is continuous or discontinuous under the present
data, the test of which would require much larger system sizes. 
\begin{figure}[t]
\centerline{\epsfxsize=9cm \epsfbox{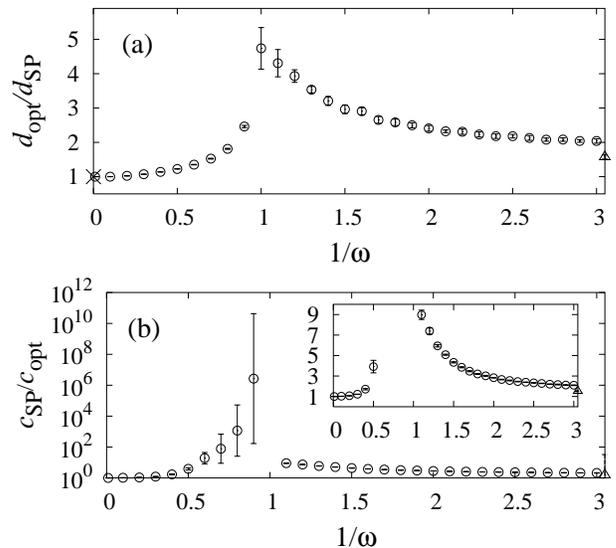}}
\caption{ 
The relative optimal path length $d_{\rm opt}/d_{\rm SP}$ (a) 
and the relative advantage $c_{\rm SP}/c_{\rm opt}$ (b) 
as a function of the exponent $\omega$ of the
cost distribution for SF network with
$\gamma=3$ and $N=10^4$. The effect of the disorder becomes
maximal at $\omega=1$, which manifests itself as the peak in both
quantities. The cross (triangle) symbol represents the
the shortest path length of the underlying binary
network (the uniform cost distribution case). 
The inset of (b) shows the same data in linear scale,
excluding those with $c_{\rm SP}/c_{\rm opt}\ge 10^1$.}
\end{figure}

As the optimal path length increases, the total load of the system
grows, as it satisfies the sum rule $\sum_i \ell_i = N(N-1)(d_{\rm opt}+1)$.
That is, by taking the optimal paths, 
we need more resource---router capacity, for example--- 
to maintain the system in the free-flow state, 
where packets can travel without congestion.
The advantage in taking the optimal paths then should compensate the 
increase of resource requirement, for it to be {\it optimal.}
We show in Fig.~1(b) the advantage in taking the optimal paths,
by the ratio of $c_{\rm SP}$, the average cost along the shortest paths,
to the $c_{\rm opt}$, that for the optimal paths.
As anticipated, the advantage is always larger than $1$,
and it is more advantageous to take the optimal paths as the disorder
becomes strong, exhibiting a strong peak near $\omega=1$.
Furthermore, $c_{\rm SP}/c_{\rm opt} > d_{\rm opt}/d_{\rm SP}$ in all cases studied, that is,
the increase in the optimal path length (the required resource)
is compensated by the cost advantage.

In the following, we focus on the three specific cases of the
cost distribution; (i) the exponential cost distribution
($\omega\to\infty$), (ii) the uniform cost distribution
($\omega\to0$), and (iii) the strong disorder limit ($\omega=1$).

\begin{figure}[t]
\centerline{\epsfxsize=9cm \epsfbox{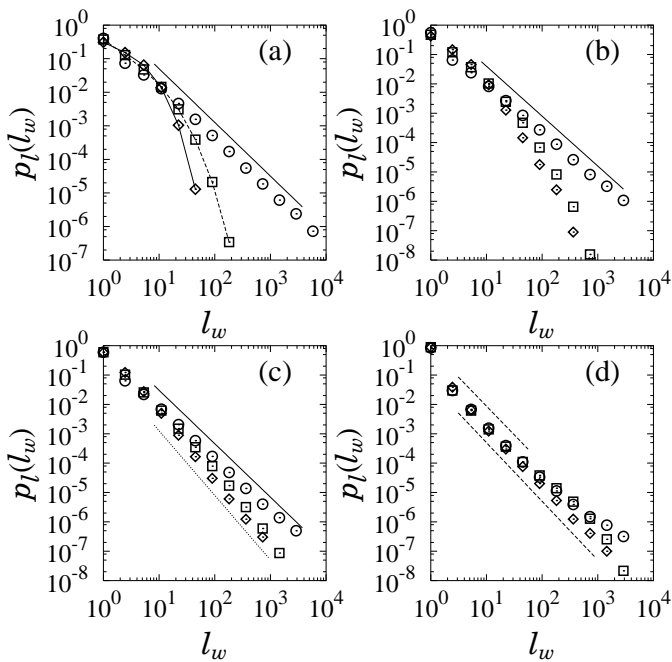}}
\caption{The load distribution of the weighted networks in the
presence of disorder. Plotted are in each panel for the ER networks
(a), the SF networks with $\gamma=4$ (b), $\gamma=3$ (c), and $\gamma=2.0$ (d). 
All networks are of size $N=10^4$. Data are logarithmically binned.
The symbols are for the exponential cost distribution 
$(\diamond)$, the uniform cost distribution $(\Box)$, and 
the strong disorder limit $(\bigcirc)$. 
The guidelines are plotted together, with the slope $-1.6$ (a),
$-1.7$ (b), $-1.8$ (solid) and $-2.2$ (dotted) (c), and $-2.0$ (d).
} 
\label{fig:weak}
\end{figure}
{\em Load distribution}--- First, in the most global level, we
study the load distribution of the networks in the presence of
the disorder. One may expect that the change in the load
distribution would be largest for the strong disorder, moderate
but substantial for the uniform cost distribution, and minimal for
the exponential distribution, as was the case for the change in
the optimal path length. This picture is confirmed by 
numerical simulations, shown in Fig.~2, for the ER network, the SF
networks with $\gamma=4$, $3$, and $2$. The load of the
weighted networks can be efficiently computed by the modified
Dijkstra algorithm~\cite{algorithm,brandes}. 
For the strong disorder limit ($\omega=1$), the equivalent
calculation can be done by noting that the optimal paths in this
case lie on the minimum spanning tree (MST) \cite{braunstein},
constructed by removing links in the descending order of their
costs one by one unless such a removal disconnects the graph. 
We exploit this fact and compute the load in the strong disorder
limit by constructing MST via, e.g., the Kruskal algorithm
\cite{algorithm}. There one can see that the load distribution
becomes broader as the strength of the disorder increases.

A surprising result is that, in the strong disorder limit, the load 
distribution follows a power law even for the ER network [Fig~2(a)]. 
Recall that the ER network has the exponentially decaying load 
distribution in its binary version. The load exponent is measured to be 
$\delta \approx 1.59(4)$ for the ER network. 
For the SF networks, it is measured that $\delta=1.64(4)$, $1,69(2)$, 
$1.73(3)$, $1.82(8)$, $1.95(5)$ and $1.96(5)$ for $\gamma=5.0$, 
$4.0$, $3.5$, $3.0$, $2.5$ and $2.0$, respectively. 
In the strong disorder limit, the MST can be regarded as the composition 
of the percolation clusters at the percolation threshold and 
inter-links, so called hot bonds, connecting disconnected 
clusters~\cite{braunstein}. 
This picture can be applied for the case of $\gamma > 3$, where 
the percolation threshold is finite.
The degree of each vertex in the MST is proportional 
to that in the original network \cite{szabo}. For example, 
the degree distribution of the MST of the ER network is not scale-free. 
In this case, it is measured that the load exponent depends on 
the degree exponent. When $2 < \gamma < 3$, however, due to 
the vanishing percolation threshold~\cite{sf-percol} and the presence 
of non-negligible fraction of short cycles \cite{lee}, 
the percolation cluster network picture no longer holds. 
The formation of the MST is not random and the 
proportionality of the degrees of the vertex in the MST 
and in the original network breaks down. 
In this case we find numerically that the load distribution 
follows a power law with the exponent about $2$
with an additional fatter tail.

In the weak disorder limit, the sum of all the costs along the path
determines the optimal path. The load distribution depends on
the strength of disorder. Note that in this case,
the load distribution for the ER network does not follow
a power law. Numerical data for other values of $\gamma$
are also shown in Fig.~2.

{\em Load-degree scaling}--- Next, to probe loads in the more microscopic
settings, we focus on how the load of an individual vertex would change 
by the presence of disorder. In the binary networks, there is a scaling 
relation between the degree and the load of a vertex, as 
\be 
\ell_b \sim k^{\eta}, 
\label{relation}
\ee 
with
$\eta=(\gamma-1)/(\delta-1)$~\cite{load}, where the subscript 
$b$ denotes the binary network. 
When the disorder becomes sufficiently strong, however,
it is not clear if such a scaling relation would still hold. We
find indeed there exists strong dispersion in the scaling relation
for the uniform cost distribution. To characterize the
dispersion, we consider the conditional probability $p(\ell_w|k)$
that the load of a vertex with degree $k$ is $\ell_w$. If this
distribution is sufficiently broad, it is meaningless to speak of
a scaling relation as in the binary network version. We show the
result of numerical simulation specifically for $k=2$ in Fig.~3.
In the strong disorder limit, $p(\ell_w|k)$ even
follows the similar power-law decay as $p_{\ell}(\ell_w)$, meaning that
the degree of a vertex has essentially no correlations with
the load. This picture also applies to the SF network, as shown
in Fig.~3(b) for the case of $\gamma=3$. 
We also find that it is also the case for intermediate degree nodes, 
$k=10$ for example, in the SF networks with $\gamma=3$.
Thus in the presence of disorder, 
one may not predict the level of traffic at a router or
the centrality of an individual based solely on the connectivity
information.
\begin{figure}
\centerline{\epsfxsize=9cm \epsfbox{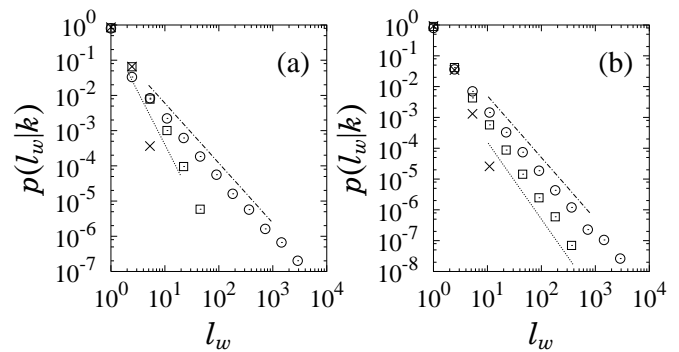}}
\caption{The condition probability $p(\ell|k)$ with $k=2$ for the
ER network (a) and for the SF network with $\gamma=3$ (b). Data
are for the exponential cost distribution ($\times$),
the uniform cost distribution ($\Box$), 
and the strong disorder limit ($\bigcirc$).
Plotted are the guidelines with slopes $-3$ and $-1.7$ in (a);
$-2.5$ and $-2$ in (b).
} \label{dispersion}
\end{figure}

{\em Disorder vs.\ Network heterogeneity}--- We now turn our
attention to the interplay between the disorder in edge costs and
the network heterogeneity. To characterize the extent of the
effect due to the disorder by a simple scalar measure, we
introduce the Pearson correlation coefficients between the degree 
($k$) and the load ($\ell_w$) in the weighted network, and 
between $\ell_w$ and the load ($\ell_b$) in the binary network, 
of the same vertex. The correlation coefficient is defined by 
$r_{xy}\equiv {(\overline{xy}-\overline{x}\cdot\overline{y})}/{\sigma_x\sigma_y}$,
where 
$\overline{x}$ and
$\sigma_x$ denote the average and the standard deviation, 
respectively, of a variable $x$ over all vertices.
($x,y$) stands for ($k,\ell_w$) or ($\ell_w, \ell_b$).

\begin{figure}
\centerline{\epsfxsize=9cm \epsfbox{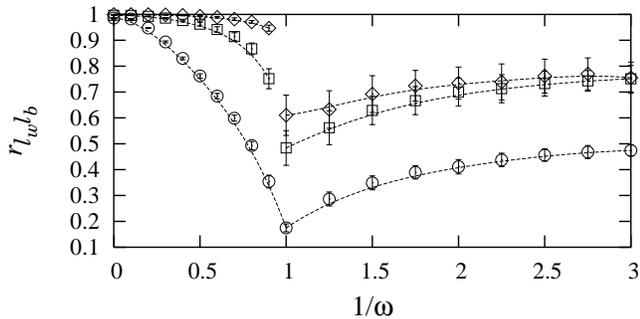}}
\caption{The correlation coefficient $r_{\ell_w\ell_b}$
as a function of $\omega$ for the ER network ($\circ$),
the SF network with $\gamma=3$ ($\Box$) and $\gamma=2$ ($\diamond$).
The dotted lines interpolating the data are drawn for the eye.
The plot of $r_{k\ell_w}$, which is not shown here, follows 
very similar feature, owing to the high correlation between $k$ and $\ell_b$.
}
\end{figure}
As the scaling relation Eq.~(\ref{relation}) implies, the correlation
$r_{k\ell}$ for binary network is high, typically larger than 0.9.
This correlation decreases as the strength of disorder
increases and the effect of disorder is larger for the ER network than for
the SF network.

The result that the effect of disorder is larger for the ER
network than for the SF network may be understood as 
follows: The binary ER network is homogeneous, so are
the pathways inside it. As we turn on the disorder, this
homogeneity breaks down and the extent to which this induced
heterogeneity becomes larger as the strength of disorder
increases. Due to this induced heterogeneity, {\em de novo}
hierarchy in vertices \cite{hierarchy} builds up and according to
this hierarchy, the pathway structure of the weighted ER network are
rearranged. In SF networks, however, such heterogeneity and
hierarchy in vertices exist even in the absence of disorder.
Furthermore, such an inherent heterogeneity suppresses the effect of
disorder in a way that the larger the degree of a vertex is, the
more likely it is to take an edge with very small cost. As a
result, the disorder in edge cost competes with the heterogeneity
of the network to give a full effect, and thus the effect of
disorder becomes weaker as the heterogeneity of a network
increases ($\gamma$ decreases), as can be seen in Fig.~4.

To summarize, we have studied the optimal transport in 
weighted complex networks by extending the notion of the load 
used in binary networks to weighted networks and investigated 
how the transport property is changed accordingly.
We found that the load distribution in the strong disorder limit
is related to the structure of the minimum spanning tree.
The load distribution follows a power law even for the ER 
network, and the load exponent $\delta$ depends on $\gamma$ 
for the SF network with $\gamma > 3$.  
For $2 < \gamma <3$, however, $\delta\approx2$ but with additional 
fatter tail.
In the weak disorder regime, it is found that for sufficiently 
narrow cost distribution the load distribution changes little. 
As the heterogeneity of weights increases, however, the effect 
becomes significant and becomes maximal as it approaches 
the cost distribution $p_c(c)\sim
1/c$, being equivalent to the strong disorder limit. 
This situation also holds on the individual vertex level,
in that the fluctuation of the load of the vertices of a given
degree grows unboundedly as the cost distribution becomes broader.
The effect of disorder is manifestly larger for the ER networks
than in the SF networks, since the disorder must compete with the
network heterogeneity in SF networks.
Finally, we note that the time delay effect by packets travelling 
with constant speed does not change the load exponent even in 
weighted networks, which is checked through the weighted ER network. 

In this Letter, we have considered the uncorrelated cost
distributions on uncorrelated networks only. In the real-world as 
well as the model evolving networks, however, the weight of a link 
and the degrees of the vertices at each end are often correlated
\cite{barrat03}. The effect of such correlated disorder on the 
transport property in the weighted networks is an open question.

\begin{acknowledgments}
This work is supported by the KOSEF Grant No. R14-2002-059-01000-0
in the ABRL program.
\end{acknowledgments}

\end{document}